\documentclass[fleqn,twocolumn]{article}

\usepackage[cp1251]{inputenc}
\usepackage[english]{babel}

\usepackage{epsf}
\usepackage{amssymb}
\usepackage{amsmath}

\begin{document}

\renewcommand{\refname}{\normalsize } 
\renewcommand{\abstractname}{\normalsize \textbf Abstract}

\begin{flushleft}
UDC 530.12; 531.51%
\end{flushleft}

\begin{center}
\textbf{ON POSSIBLE MANIFESTATION OF FEEDBACK COUPLING BETWEEN GEOMETRY
AND MATTER IN A PHENOMENON OF AN ACCELERATING EXPANSION OF THE
UNIVERSE}\\[0.5cm]
\textsf{V.V. Kuzmichev, V.E. Kuzmichev}\\[0.5cm]
\textit{Bogolyubov Institute for Theoretical Physics, \\ Nat.
Acad. of Sci. of Ukraine}\\
\textit{Metrolohichna Str. 14B, 03143 Kiev, Ukraine}\\
\end{center}

\begin{abstract}
It is shown that an accelerating
expansion of the present-day Universe extracted from observed
luminosity of the type Ia supernovae can be explained by quantum
theory which takes into account feedback coupling between geometry
and matter (like in Mach's principle). At the same time an
accelerating expansion of the Universe is explained by the
influence of small negative cosmological constant. A comparison
with the model with positive cosmological constant (dark energy)
which also has obtained its theoretical grounds in a structure of
the developed formalism is made. Parameters of the Universe in the
states with large quantum numbers are calculated.
\end{abstract}

\ \\

\begin{center}
\textbf{1. INTRODUCTION}
\end{center}

An analysis of possible reasons for the observed weak 
luminosities of the type Ia supernovae (SNe Ia) at cosmological
redshift $z \approx 0.5$ \cite{R98,P99} demonstrates that this 
phenomenon cannot be put down to nonstandard evolution of their 
luminosity, absorption effects of an interstellar dust, gravitational
lensing and other physical processes which are not connected with the
overall expansion of the Universe as a whole (see discussion in 
\cite{F03,T03,R04}).  In accordance with the principles of general 
relativity the observed dimming of the SNe Ia can be interpreted
as an evidence of an accelerating expansion of the present-day Universe.
Phenomenological models which are used herewith suppose an existence of
nonzero cosmological constant in the Universe treated as vacuum energy density
or hypothetical cosmological liquid with negative pressure (so-called dark energy
\cite{O95,B99}). The models of such type allow to describe the available 
dataset on the distance moduli of the SNe Ia depending on redshift 
by fixing free parameters (e.g. from a $\chi^{2}$ statistic). 

Providing a formal agreement with modern astrophysical observational data
(SCP \cite{P99}, HST \cite{T03} and WMAP \cite{S03} projects) phenomenological 
models come across difficulties in questions of principle when trying to find
a theoretical explanation for the values of their own free parameters and their
physical motivation. Among the fundamental problems available here it is 
possible to pick out the cosmological constant problem, a task to determine
the nature of dark energy and a puzzle concerning the coincidence between the 
contributions from dark energy $\Omega_{X} \approx 0.7$ and dark matter 
$\Omega_{M} \approx 0.3$ to the total energy density nowadays 
\cite{T03,R04,PR03,C03}. It is assumed that in order to solve them 
one should exceed the limits of modern cosmology built on the
principles of general relativity \cite{C03,CD03,D02}.

In the present article the problem of an accelerating
expansion of the Universe is analysed within the framework
of cosmological quantum model \cite{V98,V99a,V99b,VV02}. The
main feature of this approach lies in taking into account
possible feedback coupling between geometry
and matter. This coupling should be taken into consideration
when one studies the processes in which the Universe appears as a whole
(on the scales that exceed significantly the size of the superclusters
of galaxies, $> 200$ Mpc).

Quantum model of the Universe characterized by nonzero
vacuum energy density, $\rho_{vac} \neq 0$, which takes into
account feedback coupling between geometry
and matter, allows to describe numerically the observed dependence
of the distance moduli of the SNe Ia on $z$ in the whole range of 
redshift measured values \cite{R04} with the same accuracy that
is achieved within the limits of the phenomenological model
with positive cosmological constant in classical cosmology.
The latter also receives its theoretical grounds in the structure 
of developed formalism.

\begin{center}
\textbf{2. Quantum geometrodynamics in the minisuperspace model}\\[0.3cm]
\textit{2.1 Motivation}
\end{center}

An available current experimental dataset allows to state
that quantum theory describes adequately properties of 
various physical systems. The universal validity of quantum theory
demands that the Universe as a whole must obey quantum laws too. The
quest for these laws falls into the realm of research of quantum 
cosmology. Since gravity dominates on cosmological (very large) scales
any consistent formalism of quantum cosmology must contain quantum 
theory of gravity. Driving forces that give reasons for quantum
gravity research are not restricted to aspirations to obtain
a unified theory of all interactions, search for mathematical
consistency, or determination of origin and nature of space and time
(a review of motivations from different points of view on the problem
one can find, e.g., in \cite{Ish95}). There exist the problems 
which remain unsolved in standard model of the hot Universe and
which, as it seems today, cannot be solved without appeal to
quantum cosmology.

It is generally accepted that the early stage of an exponential 
expansion of the Universe within the framework of inflationary
scenario of classical cosmology withdraws the horizon problem,
directs the density parameter $\Omega$ to unity and explains
the absence of registration acts of monopoles, topological defects
and etc. by very low density of these relics. But a number of
problems remains outside the limits of inflationary model. Among 
them there are the mystery of origin of primordial fluctuations of energy
density, explanation of time arrow and determination of initial
conditions of the evolution of the Universe \cite{Wil01}. Besides this,
it is established nowadays that certain problems are solved by
inflationary model in unproper way or these problems can be avoided at all
or solved differently \cite{VV02,Cou04}. For instance, the horizon
problem is in fact directly connected with the physical processes
in the Planck era \cite{Pad93} and therefore one should appeal 
to quantum gravity in order to solve it. The flatness problem
can obtain its solution within quantum description in Planck time
as well \cite{NP83}.

Inflationary scenario does not allow to tackle the problem of
presence of singularities in quantum cosmology \cite{Cou04}. 
Inflation cannot be continued infinitely into the past, mainly because
the flat de Sitter metric becomes geodesically incomplete then \cite{HE77}.
The main achievement of inflationary model is, as it is widely
accepted, an opportunity to obtain the Universe with current parameters
(such as size, energy density contrast, age etc.) starting from the natural
Planck values for different quantities (so-called \textit{small bang}).
However, the same result can be achieved in quantum cosmology as well 
\cite{V98,V99a}, which does not contradict with inflationary paradigm at 
this point. Let us note that inflationary model itself needs quantum 
cosmology for its motivation, namely in order to ensure the long enough 
duration of the inflation period (determined by numerical coefficient 
in exponent) which would agree with observations \cite{Wil01}.

Since cosmology considers the Universe as a system with very large
scales, while quantum phenomena are typical for microscopic systems,
then a combination of words \textit{quantum} and \textit{cosmology}
may seem contradictory. It is commonly accepted that quantum effects of
gravity take place at Planck length scales  $L_{P} = \sqrt{G
\hbar/c^{3}} \sim 10^{-33}$ cm, while the space-time structure 
at large scales will be classical automatically. Such point of view
is motivated only if a consistent quantum theory 
of gravity exists within whose framework the small parameter $L_{P}/L$, 
where $L$ is some typical length in the Universe, can be introduced and
an appropriate perturbation theory can be constructed, and from the latter 
it follows that quantum effects are negligibly small for $L_{P}/L \ll 1$
\cite{Ish95}. But an assumption that quantum effects of gravity are small
neglects completely the possibility of their non-perturbative character,
while if one takes it into account that could provide in particular 
quantum field theory with an appropriate cut-off.

But it should be noticed that quantum effects are not a priori 
restricted to certain scales. Rather the processes of decoherence 
(when the coherent superposition of the states turns into incoherent
one; interference effects are absent) through the environment 
can explain why quantum effects are negligible or
important for an object under consideration \cite{KJ98,Ki99,Zeh99}.

Since nowadays a consistent quantum theory of gravity has not been
formulated, a research here is conducted in a few directions. The method
of canonical quantization of constraint systems proposed by Dirac \cite{D68}
provides a basis for the theory developed in this article.

The structure of constraints which describe the evolution of intrinsic
geometry and extrinsic curvature of spacelike hypersurface in 
space-time is such that true dynamical degrees of freedom cannot 
be distinguished explicitly from quantities which determine 
hypersurface. This leads to famous problems in interpretation
of quantum geometrodynamics constructed on the basis of the Wheeler–DeWitt
equation \cite{W67}. The main reason for these difficulties is that there 
is no predetermined way to identify spacetime events in generally covariant theories
(i.e. one cannot measure the metric, but only the geometry).

In order to solve the problem mentioned above an approach related to the notion of a
medium which determines the reference frame\footnote{Application of 
\textit{material reference frames} has a long history. They were used already by 
Einstein \cite{E17} and Hilbert \cite{H17}, but in somewhat idealized form which
did not take into account a back action of material reference frames
on geometry.} (so-called \textit{reference fluid}) seems 
promising \cite{KT91,BK95,BM96,ShS04}. The problem here lies in finding an
appropriate medium (an additional source in the Einstein equations) which, when
quantizing in Dirac's formalism, would lead to functional Schr\"{o}dinger type equation.
Variables which describe a medium (the reference frame is considered as
a dynamical system) mark spacetime events. They play the role of the canonical 
coordinates which determine \textit{embedding} in surrounding spacetime,
while new constraints turn out to be linear with respect to momenta
canonically conjugate with the medium variables. Such an additional source
is introduced in the action and determines in particular the time variable.
The invariance of action here remains unbroken.  

The replacement of the Wheeler–DeWitt equation by the functional 
Schr\"{o}dinger type equation allows to introduce the positive-definite 
conserved inner product and to advance essentially in constructing of
consistent quantum theory of gravity. But a discovery of corresponding (physical)
medium, which defines the reference frame, is a nontrivial task in itself. In
Refs. \cite{V98,V99a,V99b,VV02} this problem was solved in terms of the 
minisuperspace model. We shall consider this case in more detail.

\begin{center}
\textit{2.2 Main equations}
\end{center}

Just the same as in ordinary nonrelativistic and relativistic
theories it is possible to assume that the problem of evolution and research
into the properties of the Universe as a whole can be reduced to solution
of a partial differential equation which determines eigenvalues and
eigenfunctions of some Hamiltonian-like operator (in the space of
generalized variables whose roles are played by metric tensor components
and matter fields). For simplicity we restrict our study to the case of minimal
coupling between geometry and matter. Taking into account that scalar fields
play fundamental roles both in quantum field theory (see, e.g., \cite{PS01})
and in cosmology of the early Universe \cite{D88,L90,LR99}, we assume that
the Universe is filled \textit{ab initio} by primordial matter in the form of
a scalar field $\phi$ with some potential energy density $V(\phi)$.

We suppose that the Universe as a whole is homogeneous, isotropic and
spatially flat, and a scalar field $\phi$ is uniform. The geometry of
such a Universe is determined by the famous Robertson–Walker metric \cite{W72}.
From the principle of least action it follows the constraint equation
\cite{V98,VV02,V04}, $\delta S/\delta N = 0$, where $S$ is a corresponding
action functional, and $N$ is a lapse function that specifies the time
reference scale and plays the role of a Lagrange multiplier in the
ADM formalism \cite{ADM}, which is the Einstein-Friedmann equation for
the $(^{0}_{0})$ component.

The structure of the constraint is such that true
dynamical degrees of freedom cannot be singled out explicitly.
In the model considered, this difficulty is reflected
in that the choice of the time variable is ambiguous (so-called
\textit{problem of time}). In order to solve this problem
in general relativity it will be enough to supplement the field
equations with a coordinate condition which does not change
the Einstein equations themselves, but only specifies the
spacetime platform from which one observes the gravitational field
(the enlarged system of constraints is no longer first class
and it is possible to eliminate \textit{non-dynamical} variables).
But this method does not allow to solve the problem of time
for a quantum description \cite{KT91}.

Therefore we shall use another approach in which a coordinate 
condition is imposed prior to varying the action functional
and included in it with the aid of a Lagrange multiplier.
Parametrization of the action functional (see, e.g., 
\cite{D68,KT91,K76}) restores its coordinate invariance
expressing it in arbitrary coordinates. At the same time the
\textit{privileged} time coordinate introduced by means of
the coordinate condition is adjoined to the field variables
and takes the role of the medium variable which determines 
the reference frame.

We will choose the coordinate condition in the form
\begin{equation}\label{01}
    T' = N,
\end{equation}
where $T$ is the new field variable (the privileged time coordinate),
while differentiation with respect to arbitrary variable (conformal
time or \textit{arc parameter}) $\eta$ is denoted by a prime, the 
parameter $\eta$ is related to the synchronous
proper time $t$ by the differential equation $dt = N a d \eta$, $a$
is a cosmological scale factor.

We shall include the coordinate condition (\ref{01}) in the action 
functional with the aid of a Lagrange multiplier $P$ and obtain
the modified action of the minisuperspace model in the conventional form
\begin{equation}\label{02}
  S_{mod} = \int \! d\eta\, \left[\,\pi _{a}\,a' + \pi _{\phi }\,
  \phi' + P\,T' - N\,\mathrm{H}\,\right],
\end{equation}
where $\pi _{a}$ and $\pi _{\phi }$ are the
momenta canonically conjugate with the variables $a$ and $\phi$, and
\begin{equation}\label{03}
    \mathrm{H} = \frac{1}{2}\,\left( -\,\pi _{a}^{2}
       + \frac{2}{a^{2}}\,\pi _{\phi }^{2} - a^{2}
       + a^{4}\,V(\phi ) \right) + P
\end{equation}
is the Hamiltonian to within multiplier $N$. 
Here and below we give all relations between
dimensionless units. The length is taken in units of the modified 
Planck length $l_{P} = \sqrt{2 G \hbar/(3 \pi c^{3})} = 0.744 \times 10^{-33}$ cm, 
the energy density is measured in units of 
$\rho_{P} = 3 c^{4}/(8 \pi G l_{P}^{2}) = 1.627 \times
10^{117}\, \mbox{GeV cm}^{-3}$ and so on.

The variation of the action (\ref{02}) with respect to $N$ leads to
the constraint equation
\begin{equation}\label{04}
    \delta S_{mod}/\delta N = 0 \quad \Rightarrow \quad \mathrm{H} = 0.
\end{equation}
The parameter $T$ can be used as an independent variable
for the description of the evolution of the Universe both in
classical and quantum cosmology \cite{VV02}.

In quantum theory the constraint equation (\ref{04}), in accordance
with a procedure proposed by Dirac \cite{D68},  comes to
be a constraint on the wavefunction $\Psi$,
\begin{equation}\label{05}
    i\,\partial_{T} \Psi = \hat{\mathcal{H}} \Psi,
\end{equation}
with a Hamiltonian-like operator
\begin{equation}\label{06}
    \hat{\mathcal{H}} = \frac{1}{2} \left(\partial_{a}^{2} -
    \frac{2}{a^{2}}\,\partial_{\phi}^{2} - a^{2} + a^{4} V(\phi)
    \right ),
\end{equation}
where we have introduced the operators $P = -i\,\partial_{T}$, $\pi _{a} =
-i\,\partial_{a}$ and $\pi _{\phi } = -i\,\partial_{\phi }$, which
satisfy the ordinary canonical commutation relations, $[T,P] = i$, $[a,\pi
_{a}] = i$, $[\phi, \pi _{\phi }] = i$, while others vanish.

The wavefunction $\Psi$ depends on a cosmological scale
factor $a$, a scalar field $\phi$ and time coordinate $T$.
One can introduce, at least formally, a positive definite scalar product
$\langle \Psi|\Psi \rangle < \infty$ and specify the norm of a state.
This makes it possible to define a Hilbert space of physical states and
to construct quantum mechanics for the model of the Universe
being considered.

Equation (\ref{05}) has a particular solution with separable variables
\begin{equation}\label{07}
    \Psi = \mbox{e}^{\frac{i}{2} E T} \psi_{E},
\end{equation}
where the wavefunction $\psi_{E}$ satisfies the time-independent equation 
\begin{equation}\label{1}
 \left( -\,\partial _{a}^{2} + \frac{2}{a^{2}}\,\partial _{\phi }^{2} + U - E  \right)
 \psi _{E} = 0,
\end{equation}
and
\begin{equation}\label{2}
U = a^{2} - a^{4} V(\phi)
\end{equation}
can be interpreted as an effective potential.

The function $\psi _{E}$ is specified in space of two variables, $a$ and $\phi$.
In classical approximation the eigenvalue $E$ determines the components of 
the energy-momentum tensor
\[
  \widetilde T^{0}_{0} = \frac{E}{a^{4}},\quad
 \widetilde T^{1}_{1} = \widetilde T^{2}_{2} = \widetilde T^{3}_{3} =
 -\,\frac{E}{3\, a^{4}},
\]
 \begin{equation}\label{08}
 \widetilde T^{\mu }_{\nu } = 0
  \ \ \mbox{for} \ \  \mu \neq \nu,
\end{equation}
which in the case $E > 0$ describes an additional source of the
gravitational field in the form of relativistic matter of an arbitrary nature. 
Equation (\ref{1}) formally turns into the Wheeler–DeWitt
equation for the minisuperspace model \cite{W67} in the
special case $E \rightarrow 0$.

\ \\

\begin{center}
\textit{2.3 Model of a scalar field}
\end{center}

The quantum state $\psi _{E}$ depends on the form and numerical
value of $V(\phi)$. We shall use the model of a scalar field which
slowly (in comparison with rapid motion with respect to
the variable $a$) rolls from some initial value $\phi_{start}$ with
the Planck energy density $V(\phi_{start}) \sim 1
$\footnote{The evolution of the Universe in time can be considered in accordance with 
classical conceptions starting from this value of the energy density \cite{L90}.}
to the equilibrium state $\phi_{vac}$ with the energy density
$\rho_{vac} = V(\phi_{vac}) \ll 1$. This constant density determines
a cosmological constant $\Lambda = 3 \rho_{vac}$. At the next stage
of the evolution the scalar field oscillates with a small amplitude
near $\phi_{vac}$ under the action of quantum fluctuations. In such
a model the motion with respect to $\phi$ always will be finite.

The analogous model of the scalar field was considered for the first time
in connection with inflationary scenario (see, e.g., \cite{L90,LR99} and
references therein)\footnote{We shall note that in light of the
coincidence problem (the contributions from dark energy and dark 
matter to the total energy density in the Universe have the same order of magnitude),
the model of \textit{quintessence} - it is a scalar field $\varphi$ of special type
with potential energy density $\mathcal{V}(\varphi)$ modeled by
different functions (see, e.g., review \cite{PR03}) - 
is widely discussed in the literature. There is a fundamental difference
between the quintessence $\varphi$ and the scalar field $\phi$ of the
model being considered in this article. The field $\phi$ models
primordial matter which is a source of real matter \cite{VV04b},
including the quintessence $\varphi$ (if it really exists). The question
concerning the relation between the fields $\phi$ and $\varphi$ goes 
beyond the scope of this article and will not be considered further.}.
For inflationary model the presence of minimum in the function
$V(\phi)$ is of great importance. The oscillations of the scalar field
near a state of equilibrium with subsequent transfer of energy of 
these oscillations to \textit{real} particles allow to fill
the Universe, which has become empty after the exponential expansion, 
with hot matter \cite{L90,Kof96}.

\begin{center}
\textit{2.4 Solution of the time-independent equation}
\end{center}

For positive definite function $V(\phi)$ an effective potential
$U$ as a function of $a$ has a form of barrier. In this case
the Universe described by equation (\ref{1}) can be both in
continuum states with $E > 0$ and quasistationary ones which
correspond to complex values $E = E_{n} + i\,\Gamma_{n}$,
where $E_{n} > 0,\ \Gamma_{n} > 0$ and $\Gamma_{n} \ll E_{n}$
\cite{V99a,VV02}.  Quasistationary
states are the most interesting since the Universe in such states
can be described by the set of standard cosmological parameters
(Hubble constant, deceleration parameter, mean energy density,
density contrast, amplitude of fluctuations of the cosmic microwave background 
radiation temperature and so on) \cite{VV02}.

Taking into account that a motion with respect to $a$ in the early 
Universe is supposed to be rapid in comparison with the slow
variation of the state of the scalar field we find that the wavefunction
$\psi _{E}$ of quasistationary state, considered as a function of $a$
at fixed field $\phi$, has a sharp peak and it is concentrated mainly
in the region limited by the barrier (\ref{2}) \cite{VV02,V04}.
Then following Fock \cite{F76} one can introduce an approximate 
function which is equal to exact wavefunction inside
the barrier and vanishes outside it. Taking into account
finite motion with respect to $\phi$, this function can be
normalized and used in calculations of expectation values of observed
parameters. Such an approximation does not take into account exponentially small
probability of tunneling through the barrier $U$ in the region of large 
values of $a$, where $a^{2} V > 1$ \cite{V99a,VV02}. It is valid for
calculations of mean observed parameters of the Universe within its 
lifetime in given quasistationary state\footnote{At $V \lesssim 10^{-122}$ 
this time can reach the values close to the age of our Universe \cite{VV02}.}
when this state can be considered as stationary one. Here we have a close analogy
with the corresponding conclusions of ordinary quantum mechanics \cite{B71}.
In the region of large values of $a$ outside the barrier the WKB approximation
is valid \cite{V98} and the solution of equation (\ref{1}) can be 
written in an explicit form \cite{V98,V99a,VV02,V04}.

Let us consider the solution of equation (\ref{1}) in the approximation
of finite motion with respect to variables  $a$ and $\phi$. It is convenient to
expand the wavefunction $\psi _{E}$ on the basis of the functions $\langle a|n \rangle$
of oscillator 
\begin{equation}\label{5}
    \left( -\,\partial _{a}^{2} +  a^{2}  - \epsilon_{n}^{0} \right)
    |n \rangle = 0,
\end{equation}
where $a \geq 0$, $\epsilon_{n}^{0} = 4n + 3$, $n = 0,1,2, \dots$ is 
a number of state. This expansion has the form
\begin{equation}\label{6}
    \psi _{E} = \sum_{n} |n \rangle f_{n}.
\end{equation}
Functions $f_{n} (\phi)$ satisfy the set of differential equations
\begin{equation}\label{7}
    \partial_{\phi}^{2} f_{n} + \frac{1}{2} \sum_{n} K_{n n'}
    f_{n'}= 0
\end{equation}
with the kernel 
\begin{equation}\label{8}
    K_{n n'} (\phi; E) = \langle n| a^{2} |n' \rangle
    \left[\epsilon_{n}^{0}- E \right] - \langle n| a^{6} |n'
    \rangle \, V(\phi).
\end{equation}
In classical theory the gravitational field is determined
by the spacetime metric \cite{L88}. According to (\ref{5})
the states $\langle a|n \rangle$ will describe geometrical properties
of the Universe as a whole in quantum theory. A motion with respect to 
$a$ can be quantized. The correspondent equidistant spectrum of energy
has the form $\mathcal{E}_{n} = m_{P} (N + \frac{1}{2})$, where $m_{P}$ 
is the Planck mass, and $N = 2 n + 1$ gives the number of
elementary quantum excitations of the vibrations of oscillator (\ref{5}).
Mass of elementary excitations of geometry coincides with mass of known
Markov maximons which are particle-like formations with the Planck 
mass\footnote{Let us remind that the notion about massless gravitons as
gravitational field quanta was introduces within the framework of the
theory constructed in weak gravitational field approximation (gravitational
waves). It is obvious that for cosmologically significant effects this 
approximation is not valid.}.

\begin{center}
\textbf{3. Universe in the states with large quantum numbers}
\end{center}

Bearing in mind future application of the developed formalism
to the interpretation of astrophysical observational data for our Universe
we shall consider the cosmological equations obtained above in
the approximation of large quantum numbers.

Direct calculations \cite{V98,V99a} demonstrate that in the quantum
model of the Universe with the slow-roll potential energy density
$V(\phi)$ the first quasistationary state emerges when the density
reaches the value $V = 0.08$. This state is characterized by finite
values of energy density, $\rho \sim \rho_{P}$, and scalar 
curvature, $R \sim l_{P}^{-2}$, while the singular state with
$\rho \sim \infty$ and $R \sim \infty$ is excluded from consideration
as non-physical.

When $V(\phi)$ decreases to the value $V \ll 0.1$, the number
of available states of the Universe increases up to $n \gg 1$.
Before the instant when the scalar field reaches its equilibrium state
$\phi_{vac}$ the Universe may get into the state with the number
$n \gg 1$. Really, the origin of new quantum levels and the (exponential)
reduction of the width of the states that have emerged earlier lead to a
competition between the processes of tunneling through the potential 
barrier $U$ from a given $n$-th state and allowed transitions 
between the states, $n \rightarrow n \pm i$, where $i = 1,2$ \cite{V99a,VV02}.
A comparison between the probabilities of these processes demonstrates
that the process $n \rightarrow n + 1$ appears to be the most probable. Such
transitions are realized at the expense of energy of the scalar field accumulated
in the state $\phi_{start}$.

Taking into account an explicit form of the matrix elements 
$\langle n' |a^{2}| n \rangle$ and $\langle n' |a^{6}| n \rangle$,
we find that in the limiting case $n \gg 1$ the set of equations
(\ref{7}) is reduced to one equation in the approximation 
$f_{n} \approx f_{n \pm j}$, where $j = 1,2,3$. This approximation
preserves the orthogonality of the states with respect to quantum number
($s$) that characterizes the field $\phi$.

Equation for $f_{n}$ as a function of new variable $x =
\sqrt{m/2}\,(2N)^{3/4}\,(\phi - \phi_{vac})$ which describes
the deviation of the field $\phi$ from the equilibrium
value $\phi_{vac}$ has the form
\begin{equation}\label{9}
    \left[\partial_{x}^{2} + z - v(x) \right] f_{n}(x) = 0,
\end{equation}
where we denote $z = (\sqrt{2N}/m) \left(1 - E/(2N)\right)$, $v(x)
= (2N)^{3/2}\,V(\phi)/m$, and $m$ is some parameter. It is
convenient to choose 
$m^{2} = \left[\partial_{\phi}^{2} V(\phi)\right]_{\phi_{vac}} >
0$. We shall assume that the density $V(\phi)$ near the point
$\phi_{vac}$ is a smooth enough function. Then expanding
$v(x)$ into Taylor's series near the point $x = 0$, we obtain
\begin{equation}\label{10}
    v(x) = v(0) + x^{2} + \alpha\,x^{3} + \beta\,x^{4} + \dots ~,
\end{equation}
where
\[    \alpha = \frac{\sqrt{2}}{3}\,m^{-5/2}\,(2N)^{-3/4}\,
    \left[\partial_{\phi}^{3}V(\phi)\right]_{\phi_{vac}},
\]
\[   \beta = \frac{1}{6}\,m^{-3}\,(2N)^{-3/2}\,
    \left[\partial_{\phi}^{4}V(\phi)\right]_{\phi_{vac}}.
\]
Since $N \sim n \gg 1$, then $|\alpha| \ll 1$ and $|\beta| \ll 1$,
and equation (\ref{9}) with the potential (\ref{10}) can be solved
using the perturbation theory for stationary problems with a 
discrete spectrum. We take for the state of the unperturbed problem 
the state of the harmonic oscillator described by equation (\ref{9}) 
with the potential (\ref{10}) for $\alpha = \beta = 0$. As a result we obtain
\begin{equation}\label{12}
    z = 2 s + 1 + v(0) + \Delta z,
\end{equation}
where $s = 0,1,2,\dots $ is number of the states of the field $\phi$, $\Delta z$
takes into account its self-action (an explicit form of $\Delta z$ is given in
\cite{VV04b}). The spectrum of energy states of the field $\phi$ has the form
$M' = M + \Delta M$, where
\begin{equation}\label{13}
    M = m \left(s + \frac{1}{2}\right),
\end{equation}
and $\Delta M = m \Delta z/2$. One can demonstrate that the following 
estimation is valid
\begin{equation}\label{14}
\frac{\Delta M}{M} \ll 1 \quad \mbox{at} \quad s > m^{-2}.
\end{equation}
Hence it appears that at large enough values of $s$ one can neglect
the self-action of the field $\phi$. It is reasonable to interpret 
$M$ (\ref{13}) as a quantity of matter/energy in the Universe
represented in the form of a sum of elementary quantum excitations of the 
vibrations of the field $\phi$ near the equilibrium state 
$\phi_{vac}$ with the masses $m$; $s$ is the number of such excitations.
For instance, for $m \sim 1$ GeV the condition (\ref{14}) is satisfied
at $s > 10^{38}$. Assuming $s \sim 10^{80}$ (the equivalent number of baryons
in our Universe) we obtain a restriction on mass of quantum excitations 
from below, $m > 10^{-21}$ GeV.

Taking into account the relation between $z$ and $E$, from (\ref{12})
we obtain the expression for the eigenvalue
\begin{equation}\label{15}
    E = 2 N - (2 N)^{2} \rho_{vac} - 2 \sqrt{2 N} M'.
\end{equation}
The wavefunction of the Universe in the state with large quantum
numbers, $n \gg 1$, $s \gg 1$, has the form
\begin{equation}\label{16}
    \psi_{E} (a, \phi) = \varphi_{n}(a) f_{ns}(\phi),
\end{equation}
where
\begin{equation}\label{17}
    \varphi_{n}(a) = \left(\frac{4}{2N + 1}\right)^{1/4} \cos
    \left(\sqrt{2 N + 1}\, a - \frac{\pi N}{2}\right),
\end{equation}
\begin{eqnarray}\label{18}
    &&  \!\!\!\!\!\!\!\!\!\!\!\!\!\!\!\!\!
    f_{ns}(\phi) = \left(\frac{m\, (2N)^{3/2}}{2\, (2s +
    1)}\right)^{1/4} \times  \\
    &\times & \cos
    \left(\sqrt{(2 s + 1)\, \frac{m}{2}\, (2 N)^{3/2}}\, (\phi - \phi_{vac}) - \frac{\pi
    s}{2}\right). \nonumber
\end{eqnarray}
These functions are normalized to unity in the ranges
$0 \leq a \leq a_{c}$ and $\phi_{-} \leq \phi \leq \phi_{+}$
limited by the classical turning points
\[a_{c} = \sqrt{2 N + 1}, \qquad \phi_{\pm} = \phi_{vac} \pm
\left(\frac{2\, (2 s + 1)}{m\, (2 N)^{3/2}}\right)^{1/2}\] 
for corresponding oscillator potentials. Beyond these ranges
an exact wavefunction decreases exponentially. Here a perfect analogy
with the normalization of quasiclassical functions in quantum
mechanics may be observed (see, e.g., \cite{M75}).

Taking into account that the mean value of the scale factor
$\langle a \rangle$ in the state (\ref{16}) is equal to
\begin{equation}\label{19}
    \langle a \rangle = \frac{1}{2}\,\sqrt{2 N + 1},
\end{equation}
we come to a conclusion that $v(x)$ in equation (\ref{9}) is
the potential energy of the scalar field contained in the Universe
with the volume $\sim \langle a \rangle^{3}$, and the variable
$x^{2}$ characterizes the deviation squared of the field $\phi$
from an equilibrium state in such a volume.
Thus equation (\ref{9}) describes the stationary
states which characterize the scalar field $\phi$ in the Universe 
as a whole. The quantities $v(x)$, $x^{2}$ and $M$ are its overall
characteristics.

Taking (\ref{19}) into account, the condition (\ref{15}) can be rewritten
in the form of feedback coupling relation between geometrical and
energetic characteristics of the Universe
\begin{equation}\label{20}
    \langle a \rangle = M + \frac{E}{4 \langle a \rangle} + 4 \langle a \rangle^{3}
    \rho_{vac},
\end{equation}
where we discard a small addition $\Delta M$ and take into consideration
that $N \gg 1$. Here the second summand on the right describes
the energy of relativistic matter, while the third term gives the
contribution from the vacuum of the scalar field.

Equation (\ref{20}) can be interpreted as one of possible implementations
of famous Mach's principle \cite{D64}. Indeed, passing to dimensional
quantities we obtain
\[\frac{G}{c^{2}} \frac{\mathcal{M}}{\mathcal{R}} \sim 1, \]
where $\mathcal{M}$ and $\mathcal{R}$ are measures of mass (without
taking gravitational interaction between bodies into account) and 
radius of the observed part of the Universe. This relation follows from
the Lense-Thirring effect in general relativity as well. In this connection
the Universe appears like a huge system which tracks and adjusts its
parameters according to feedback coupling condition (\ref{20}) 
(see also \cite{D64}).

\begin{center}
\textbf{4. Cosmological models}
\end{center}

Using the relation for mean values of a product of operators
\cite{VV04}
\[ \left \langle - \frac{1}{a^{4}} \partial_{a}^{2} \right \rangle =
\left \langle \left(\frac{1}{a} \frac{d a}{d t} \right)^{2} \right
\rangle,
\]
where $t$ is the synchronous proper time, while averaging is 
performed over the state $\psi _{E}$ normalized in a way indicated above,
from equation (\ref{1}) one can pass to the relation between
expectation values. Assuming that the mean $\langle a \rangle$
in such a state determines the scale factor in classical description
in general relativity, we obtain the Einstein-Friedmann equation
in terms of mean values
\begin{equation}\label{3}
    \left(\frac{1}{\langle a \rangle}\,\frac{d \langle a
    \rangle}{dt}\right)^{2} = \langle \rho \rangle - \frac{1}{\langle a
    \rangle^{2}},
\end{equation}
where
\begin{equation}\label{4}
    \langle \rho \rangle =
    \frac{2}{\langle a \rangle ^{6}} \left \langle -\,\partial_{\phi}^{2} \right
    \rangle + \left \langle V \right \rangle + \frac{E}{\langle a \rangle ^{4}}
\end{equation}
is the mean total energy density. In this equation the dispersion
$\sigma^{2} = \langle a^{2} \rangle - \langle a
\rangle^{2}$ and the higher-order moments with respect to deviation 
of $a$ from its mean value $\langle a \rangle$ are not taken into account.
For the problems considered in the present article they can be neglected.

The mean total energy density in the state (\ref{16}) equals to
\begin{equation}\label{21}
    \langle \rho \rangle = \gamma \frac{M}{\langle a
    \rangle^{3}} + \rho_{vac} + \frac{E}{\langle a \rangle^{4}},
\end{equation}
where $\gamma = 193/12$ is a numerical coefficient which
appears in calculation of expectation values of the operators
of the kinetic and potential parts of the energy density of
the scalar field in expression (\ref{4}). The mean density
(\ref{21}) is the sum of the energy density connected with matter
(in the form of elementary quantum excitations of the vibrations
of the scalar field near the equilibrium state $\phi_{vac}$),
the vacuum energy density and the energy density of relativistic matter.

Taking (\ref{21}) into account, equation (\ref{3}) can be rewritten
in the form of relation for the Hubble constant $H = (1/\langle a
\rangle)\,(d \langle a \rangle / dt)$ as a function of the cosmological
redshift $z = a_{0}/ \langle a \rangle - 1$,
\begin{eqnarray}\label{22}
   H^{2}(z) / H_{0}^{2} &=& \Omega_{M} (1 + z)^{3} +
    \Omega_{vac} + \\
    &+& \Omega_{R} (1 + z)^{4} + (1 - \Omega_{0}) (1 +
    z)^{2}, \nonumber
\end{eqnarray}
where
\[\Omega_{M} = \frac{\gamma M}{a_{0}^{3} H_{0}^{2}}, \quad
\Omega_{vac} = \frac{\rho_{vac}}{H_{0}^{2}}, \quad \Omega_{R} =
\frac{E}{a_{0}^{4} H_{0}^{2}}\]
are the components of the total energy density $\Omega_{0} = \Omega_{M} +
\Omega_{vac} + \Omega_{R}$ at $z = 0$, $a_{0} \equiv \langle a
\rangle _{z = 0}$, $H_{0} \equiv H(0)$. If the quantity $M$ is assumed
to be constant, then expression (\ref{22}) will describe the
evolution of the Universe in the model with a cosmological constant
(\textit{MCC}) represented in terms of mean values. If one
establishes a direct correspondence between classical values and corresponding
mean values, then such a model will be equivalent to the model
with a cosmological constant of classical cosmology \cite{PR03}.
In this case the feedback coupling between geometry and matter given
by relation (\ref{20}) is not taken into consideration.

Account for (\ref{20}) in (\ref{21}) leads to the mean energy 
density \footnote{Since relation (\ref{20}) connects overall
characteristics of the Universe, then the energy density in the
form (\ref{23}) describes only its properties as a homogeneous
system on very large scales. The density (\ref{23}), for instance,
cannot be used in calculation of fluctuations of the density near 
the mean value $\langle \rho \rangle$, which lead to formation 
of visible structures in the Universe. It is necessary to use 
the representation (\ref{21}) in order to study such processes. (See also 
\cite{C03}.)} 
\begin{equation}\label{23}
    \langle \rho \rangle = \frac{\gamma}{\langle a
    \rangle^{2}} + \widetilde{\rho}_{vac} + \widetilde{\rho}_{rad},
\end{equation}
where we denote
\[\widetilde{\rho}_{vac} = (1 - 4 \gamma)\, \rho_{vac}, \qquad
\widetilde{\rho}_{rad} = \left(1 - \frac{\gamma}{4}\right)
\frac{E}{\langle a \rangle^{4}}.\]
Dependence of the Hubble constant on $z$ in the model with
the feedback coupling (\textit{MFC}) which has no analogue
in classical cosmology  takes the form
\begin{equation}\label{24}
    H^{2}(z) / H_{0}^{2} = \widetilde{\Omega}_{M} (1 + z)^{2} +
    \widetilde{\Omega}_{vac} + \widetilde{\Omega}_{R} (1 + z)^{4},
\end{equation}
where the components with tildes
\[\widetilde{\Omega}_{M} = \frac{\gamma - 1}{a_{0}^{2} H_{0}^{2}},
\quad \widetilde{\Omega}_{vac} = (1 - 4 \gamma)\, \Omega_{vac},
\]
\[\widetilde{\Omega}_{R} = \left(1 - \frac{\gamma}{4}\right)
\Omega_{R}\]
satisfy the obvious equality 
\begin{equation}\label{25}
    \widetilde{\Omega}_{M} + \widetilde{\Omega}_{vac} +
    \widetilde{\Omega}_{R} = 1.
\end{equation}
The total energy density at $z = 0$ equals to
\begin{equation}\label{26}
    \Omega_{0} = 1 + \frac{\widetilde{\Omega}_{M}}{\gamma - 1}.
\end{equation}

Equation (\ref{3}) with the density (\ref{23}) can be integrated in
an explicit form. Neglecting the contribution from relativistic matter
we find
\begin{eqnarray}\label{30}
    \langle a \rangle &=& \frac{a_{in}}{2}\, \left(1 + \sqrt{1 +
    \zeta^{2}}\right) \left\{\mbox{e}^{\sqrt{\widetilde{\rho}_{vac}}\, \Delta
    t} - \right. \\
    &-& \left. \left(\frac{\zeta}{1 + \sqrt{1 + \zeta^{2}}}\right)^{2}
    \mbox{e}^{- \sqrt{\widetilde{\rho}_{vac}}\, \Delta t}\right\},\nonumber
\end{eqnarray}
where $\Delta t = t - t_{in}$ is time interval counted from some initial value
$t_{in}$, when the scale factor is equal to 
$a_{in} \equiv \langle a \rangle_{t = t_{in}}$; $\zeta^{2}
= (\gamma - 1)/(\widetilde{\rho}_{vac}\, a_{in}^{2})$. From this it follows that
\[\frac{\ddot{\langle a \rangle}}{\langle a \rangle} = \widetilde{\rho}_{vac}, \]
where dots denote the second derivative with respect to time $t$.

According to (\ref{30}) in the epoch, when 
$\sqrt{\widetilde{\rho}_{vac}}\,\Delta t \ll 1$, the law of evolution
of the Universe must be close to linear, $\langle a \rangle
\approx \sqrt{\gamma - 1}\,\Delta t$. If for some redshift range
$\sqrt{\widetilde{\rho}_{vac}}\,\Delta t \sim 1$, then the Universe 
during the expansion on this time interval tends on average to an exponential 
regime, namely the expansion is realized with an acceleration.

Taking into account available current astrophysical data, 
it is interesting to apply the theory developed above to
calculation of parameters of our Universe. Below we consider
the matter-dominant era, when the contribution from
$\Omega_{R} \sim 10^{-4}$ can be neglected.

\begin{center}
\textbf{5. Parameters of the Universe}\\[0.3cm]
\textit{5.1 Distance modulus of a source}
\end{center}

If one knows $H(z)$ it is possible to calculate the luminosity distance $d_{L}$
to a source with redshift $z$,
\begin{equation}\label{27}
    d_{L} = \frac{c}{H_{0}}\, \frac{1 + z}{\sqrt{\Omega_{0} -
    1}}\,
    \sin\left(\sqrt{\Omega_{0} - 1}\, H_{0} \int_{0}^{z}\!\!\frac{d
    z}{H(z)}\right)
\end{equation}
at $\Omega_{0} > 1$. In the limiting case $\Omega_{0}\rightarrow 1$
relation (\ref{27}) describes the luminosity distance in the
spatially flat Universe. Distance modulus $\mu = m - M$ (here $m$ and
$M$ are apparent and absolute magnitudes respectively)
can be calculated with the help of the equation \cite{W72}
\begin{equation}\label{28}
    \mu = 5 \lg d_{L} + 25,
\end{equation}
where $d_{L}$ is taken in units of megaparsecs.

\begin{center}
\noindent \epsfxsize=\columnwidth\epsffile{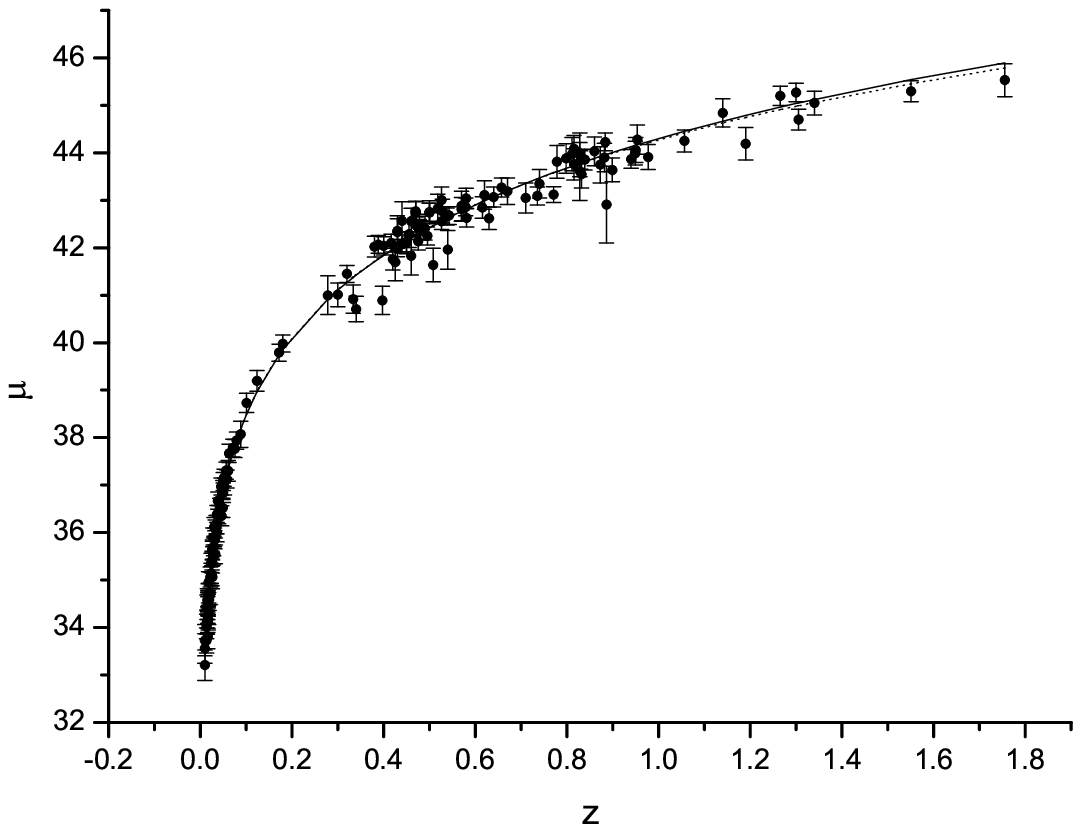}
\end{center}

\vskip-2mm\noindent{\footnotesize Fig. 1. Dependence of distance
modulus $\mu$ (\ref{28}) on redshift $z$. The result of best fitting
(according to the $\chi^{2}$ fit statistics) of quantum model (\ref{24})
(with the parameter $\Omega_{vac} = -0.0075$) using SNe Ia data (dots) 
\cite{R04} is shown as a solid line. Model with a cosmological constant 
(\ref{22}) is represented as a dotted line (for a flat Universe 
with $\Omega_{vac} = 0.71$).}%
\vskip15pt

In Fig.~1 we show the results of fitting (from a $\chi^{2}$ statistic) 
of the theoretical models (MCC and MFC) for observed distance modulus $\mu$ 
as a function of $z$ for 156 type Ia supernovae 
\cite{R04} with ``high-confidence'' spectroscopic and photometric record
for individual source (\textit{gold} SNe Ia in terminology of 
\cite{R04}). For MCC (\ref{22}) (dotted curve) the best agreement
between theory and observed data at $0.0104 \leq z \leq 1.755$
is achieved for $\Omega_{vac} = 0.71$ with $h = 0.65$, $\chi^{2} =
178$, $\chi^{2}_{dof} = 1.16$. MCC has two free parameters
($\Omega_{0} = 1$ and $\Omega_{vac}$). In this part our calculation 
agrees with results represented in \cite{R04}. The value $h = 0.65$
coincides with the Hubble constant which was obtained in \cite{F03,T03,R98}
according to dataset on supernovae at $z \lesssim 0.1$\footnote{Attachment of
additional data from WMAP experiment \cite{S03} and HST project \cite{T03} 
gives $h = 0.71^{+ 0.04}_{-0.03}$. Since in this article we consider
the problem of accelerating expansion of the Universe in the light of
information on SNe Ia only, then for self-consistent calculation the value 
of the Hubble constant which corresponds to these data should be used.}. 

In the case of MFC (\ref{24}) (a solid line in Fig.~1) the 
best agreement between theory and SNe Ia observational data 
is achieved at the value $\widetilde{\Omega}_{vac} = 0.48$
which corresponds to the density parameter
\begin{equation}\label{29}
    \Omega_{vac} = - 0.0075,
\end{equation}
with $h = 0.65$, $\chi^{2} = 181$, $\chi^{2}_{dof} = 1.17$. MFC
has only one parameter ($\widetilde{\Omega}_{vac})$. Taking into
account a degree of reliability of spectroscopic and photometric
measurements of distant sources and their possible adjustment in 
the future\footnote{Some of the data for 172 SNe Ia represented in
\cite{T03} on May, 2003 are not included in the new catalog \cite{R04}
due to their low confidence with respect to one of recorded parameters
(details see in \cite{R04}).}, one can conclude that both models
describe distance modulus of SN Ia considered as a function of redshift
$z$ practically with the same accuracy. A susceptibility level of
the $\chi^{2}$ fit statistics can be judged from the following
example. For $h = 0.664$ \cite{DD04} we obtain 
$\Omega_{vac} = 0.76$ at $\chi^{2} = 184$,
$\chi^{2}_{dof} = 1.20$ for MCC and $\widetilde{\Omega}_{vac} =
0.56$ ($\Omega_{vac} = - 0.0088$) at $\chi^{2} = 189$,
$\chi^{2}_{dof} = 1.22$ for MFC. These numbers are close to mentioned above. 

\begin{center}
\textit{5.2 Energetic and geometrical scales}
\end{center}

In the case of MCC the density parameter $\Omega_{vac} < 0$ and
$|\Omega_{vac}| \ll 1$\footnote{Let us note that the idea of
occupied levels with negative energy \cite{D79} leads to negative
energy density as well \cite{ZN71}.}. This component of energy density
forms the negative cosmological constant. At (\ref{29}) it is equal to
$\Lambda = - 1.1 \times 10^{-58}\, \mbox{cm}^{-2}$. This value is in 
good agreement with the available experimental data, $|\Lambda | <
10^{-56}\, \mbox{cm}^{-2}$ \cite{PR03}.

The total energy density (\ref{26}) equals to $\Omega_{0} = 1.03$.
It means that in the redshift range under consideration the Universe
must look like spatially flat (to within $<4~\%$). The theoretical
value of the density gets within the uncertainty limits for this
value, $\Omega_{0} = 1.02 \pm 0.02$ \cite{S03}, obtained from the 
combined data of the available astronomical observations.

The scale factor in the current epoch for the obtained
value of the parameter $\widetilde{\Omega}_{M} = 0.52$
turns out to be equal to $a_{0} = 24721$ Mpc. The same value
can be obtained directly from the solution (\ref{30}). This
number is considerably larger than the correspondent Hubble
distance, $c/H_{0} = 4612$ Mpc. Such correlation between them
gives the physical reason why the density $\Omega_{0}$
is close to unity.

Feedback coupling relation allows to estimate the total
amount of matter (the sum of masses of bodies in the Universe
taken separately, without taking their gravitational
interaction into account, according to equation (\ref{13}))
in the present-day Universe. In dimensionless units this parameter
and the scale factor are quantities of the same order of magnitude,
\[ M_{0} = 1.86\, a_{0}.\]
From here we obtain $M_{0} = 9 \times 10^{57}$ g in CGS units.

\begin{center}
\textit{5.3 Time scale}
\end{center}

The time interval $\Delta t$ counted from some instant of expansion
$z = z_{in}$, taken as a reference point, to another instant
fixed by observations with $z < z_{in}$ can be determined
using the famous expression which follows from the definition of
Hubble constant,
\[ \Delta t(z) = \int _{z}^{z_{in}}\!\! \frac{dz}{(1 + z) H(z)}.\]
Assuming $z = 0$ and $z_{in} = 1.755$, which corresponds to
most distant source SN 1997ff among SNe Ia known, we obtain
$H_{0}\Delta t(0) = 0.74$ for MFC with the parameter
(\ref{29}) and $H_{0}\Delta t(0) = 0.71$ for MCC with the parameters
$\Omega_{vac} = 0.71$ and $\Omega_{0} = 1$. This leads to
practically the same time intervals, $\Delta t(0) = 11.1 \times
10^{9}$ years for MFC and $\Delta t(0) = 10.6 \times 10^{9}$ years
for MCC. Supposing that expressions (\ref{22}) and (\ref{24})
remain valid up to singular initial state with 
$z = \infty$, we receive $H_{0} t_{0} = 1.23$ for MFC and $H_{0}
t_{0} = 0.97$ for MCC, where $t_{0} = \Delta t(0)|_{z_{in} = \infty}$
is the age of the Universe, or $t_{0} = 18.5 \times 10^{9}$ years for MFC and
$t_{0} = 14.6 \times 10^{9}$ years for MCC. Since the parameters of both models
were fitted in the finite range of $z$, then these values can be used 
for illustrative purposes only. Assuming, for example, that in the range
$1.755 < z < \infty$ the Universe is described by MFC with 
$\Omega_{vac} = 0$ \cite{VV04,V04}, then in this case we have the numerical
values of dimensionless time parameter and age of the Universe equal to
$H_{0} t_{0} = 1.10$ and $t_{0} = 16.5 \times 10^{9}$ years respectively.
These numbers get within the experimental uncertainty range of correspondent
parameters, $0.72 \lesssim H_{0} t_{0} \lesssim 1.17$ and $11 \lesssim t_{0}
\times 10^{-9}\, \mbox{years}^{-1} \lesssim 17$ \cite{PR03}, obtained  
in the analysis of old stars under the assumption that stars were formed 
not earlier that $z = 6$.

MFC predicts the distance to SN 1997ff equal to $r_{0} = c \Delta t(0)
= 3396$ Mpc. This value lays between the distances $r = 3317$ Mpc
and $r = 5245$ Mpc for sources with $z = 1$ and $z = 2$ respectively
calculated in \cite{G03} for the astrophysical data in standard model
with $\Omega_{0} = 1$ and normalization $a_{0} = c/H_{0}$. Using the
known relation $r(t) = \chi a(t)$, where $r(t)$ is a distance to
a source at the instant of time $t$ \cite{L88}, we obtain the value
of the coordinate (angular distance) $\chi$ for the source SN
1997ff, $\chi = 0.137$. That is more than 20 times smaller than
the maximum possible value $\chi_{max} = \pi$.

\begin{center}
\textit{5.4 Deceleration parameter}
\end{center}

Assuming that near $z = 0$ the deceleration parameter $q(z) =
- \ddot{a}/(a H^{2}(z))$ can be approximated by the simple expression
\begin{equation}\label{32}
    q(z) = q(0) + z \left(\frac{dq}{dz} \right)_{z = 0}
\end{equation}
and determining the free parameters $q(0)$ and $(dq/dz)_{z=0}$
from a $\chi^{2}$ statistic for SNe Ia, one can come to a conclusion \cite{R04}
that the transition between the current epoch of accelerating expansion
and previous phase with cosmic deceleration may take place at $z_{t} = 0.46
\pm 0.13$, where $q(z_{t}) = 0$. At the same time $q(0)$ restored by
the \textit{gold} sample of SNe Ia lays in the range from $-1$ to $-0.5$ (at
the $68~\%$ confidence level), or from $-1.4$ to $-0.2$ (at
the $99~\%$ confidence level). MCC with the parameters $\Omega_{vac} = 0.71$ and
$\Omega_{0} = 1$ in the approximation (\ref{32}) leads to the values
$q(0) = -0.57$ and $z_{t} = 0.46$.

For MFC (\ref{24}), in approximation linear with respect to $z$
(\ref{32}) we obtain $q(0) = - 0.48$ and $z_{t} = 0.95$. In other words
both models predict an accelerating expansion of the Universe in the current epoch
and a possible deceleration at $z > 1$.

From the solution (\ref{30}) it follows that the inflationary
expansion of the Universe, theoretically, may be realized both
in the early Universe (with the large enough value of $\widetilde{\rho}_{vac}$)
and in later epoch. This conclusion agrees with the point of view which
is widespread nowadays that the present-day Universe goes through the period
of inflationary expansion again \cite{C03,Cou04}. (In MFC we have
$\sqrt{\widetilde{\rho}_{vac}}\, \Delta t(0) = 0.5$ in the range
$0.0104 \leq z \leq 1.755$, that corresponds to observed SNe Ia.)

Let us note that the linear approximation (\ref{32}) may come to an agreement with
the SNe Ia data at $w < - 0.5$ in equation of state $p = w \rho$,
where $p$ is pressure (for MCC $w = - 1$). The more refined models
which take into account a possible dependence of $w$ on $z$ lead
to nonlinear dependence of the deceleration parameter on redshift
when processing the observational data for supernovae \cite{R04}.

\begin{center}
\textbf{6. Conclusion}
\end{center}

In the present article we demonstrate that an accelerating
expansion of the Universe observed nowadays for the SNe Ia data \cite{R98,P99,R04}
may give the evidence in favour of presence of the small negative cosmological 
constant in it, $\Lambda = -1.1 \times 10^{-58}\ \mbox{cm}^{2}$,
and be the direct confirmation of the existence of the feedback coupling
between geometry and matter on the scales that exceed significantly the size of 
the superclusters of galaxies anticipated by Mach's principle \cite{D64}. 
In quantum model of the Universe this principle is not introduced from the 
outside as an additional condition. It is contained in the theory by itself
in the form of the condition on eigenvalues $E$ (\ref{15}). The parameters
calculated in accordance with quantum mechanical principles are
in good agreement with the observational data. In particular, quantum model
does not contradict with an idea of the decelerating expansion of the Universe in
the epoch $z>1$. The largest possible distance between sources 
$r_{max}  = \pi \, a_{0} = 77663$ Mpc in the Universe described by quantum theory
can be compared with the effective particle horizon $14283$ Mpc
calculated in \cite{G03} for the spatially flat Universe. 

Exceeding the bounds of the aim of the present paper we note
that quantum cosmological model allows to solve dark matter problem
and give a natural explanation for presence of one more additional 
component in the energy density in the Universe. Here matter component
of the energy density is formed as a result of a dynamic process in
which elementary quantum excitations of the vibrations of primordial
matter (the uniform scalar field in this article) decay into real
(visible and invisible) matter mainly under the action of gravity
\cite{VV04b,VV03}. These excitations themselves are uniformly
distributed in space. They practically do not interact between themselves  
and do not make clusters with real matter, i.e. they have
properties ascribed to invisible (dark) energy \cite{T03,O95,B99},
with the exception of negative pressure, perhaps. Properties of
elementary quantum excitations of the vibrations of primordial
matter allow to identify them with invisible energy for better reason
then with invisible matter. The percentage of \textit{matter} and \textit{energy}
components in the total energy density can be made consistent with
observations on the reasonable assumptions about baryon contribution
and energy released in the decay of quantum excitations \cite{VV04b}.

{}


\begin{thebibliography}{99}
\itemsep -6pt plus 1pt minus 1pt
\bibitem{R98}\textit{Riess A.G. et al.}// Astron J. - 1998. - \textbf{116}. P.1009;
astro-ph/9805201.
\bibitem{P99}\textit{Perlmutter S. et al.}// Astrophys. J. - 1999.
- \textbf{517}. - P.565; astro-ph/9812133; Int. J. Mod. Phys. -
2000. - \textbf{A15}, S1. - P.715.
\bibitem{F03}\textit{Filippenko A.V.}, in Carnegie Observatories Astrophysics Series,
Vol. 2: Measuring and Modeling the Universe,  Freedman W.L. ed. -
Cambridge, Cambridge Univ. Press, 2003; astro-ph/0307139.
\bibitem{T03}\textit{Tonry J.L. et al.}// Astrophys. J. - 2003. - \textbf{594}.
- P.1; astro-ph/0305008.
\bibitem{R04}\textit{Riess A.G. et al.}// Astrophys. J. - 2004. - \textbf{607}. - P.665;
astro-ph/0402512.
\bibitem{O95}\textit{Ostriker J.P. and Steinhardt P.J.}// Nature. -
1995. - \textbf{377}. - P.600; astro-ph/9505066.
\bibitem{B99}\textit{Bancall N.A., Ostriker J.P., Perlmutter S.,
and Steinhardt P.J.}//  Science. - 1999. - \textbf{284}. - P.1481;
astro-ph/9906463.
\bibitem{S03}\textit{Spergel D.N. et al.}// Astrophys. J.
Suppl. - 2003. - \textbf{148}. - P.175; astro-ph/0302209.
\bibitem{PR03}\textit{Peebles P.J.E. and Ratra B.}//  Rev. Mod.
Phys. - 2003. - \textbf{75}. - P.599; astro-ph/0207347.
\bibitem{C03}\textit{Carroll S.M.}, in Carnegie Observatories Astrophysics Series,
Vol. 2: Measuring and Modeling the Universe,  Freedman W.L. ed. -
Cambridge, Cambridge Univ. Press, 2003; astro-ph/0310342.
\bibitem{CD03}\textit{Carroll S.M., Duvvuri V., Trodden M., and
Turner M.S.}// Phys. Rev. - 2004. - \textbf{D70}. - P.043528;
astro-ph/0306438.
\bibitem{D02}\textit{Deffayet C., Dvali G., and Gabadadze G.}// Phys.
Rev. - 2002. - \textbf{D65}. - P.44023; astro-ph/0105068.
\bibitem{V98}\textit{Kuzmichev V.V.}// Ukr. J. Phys. - 1998. - \textbf{43}. - P.896.
\bibitem{V99a}\textit{Kuzmichev V.V.}// Phys. At. Nucl. - 1999. - \textbf{62}. - P.708;
gr-qc/0002029.
\bibitem{V99b}\textit{Kuzmichev V.V.}// Phys. At. Nucl. - 1999. - \textbf{62}. - P.1524;
gr-qc/0002030.
\bibitem{VV02}\textit{Kuzmichev V.E. and Kuzmichev V.V.}//  Eur. Phys.
J. - 2002. - \textbf{C23}. - P.337; astro-ph/0111438.
\bibitem{Ish95}\textit{Isham C.J.}, in GR14 conference. - Florence, 1995;
gr-qc/9510063.
\bibitem{Wil01}\textit{Wiltshire D.L.}, in Cosmology: The Physics of the
Universe, Robson B., Visvanathan N. and Woolcock W.S. eds. - World
Scientific, Singapore, 1996; gr-qc/0101003.
\bibitem{Cou04}\textit{Coule D.H.}// gr-qc/0412026.
\bibitem{Pad93}\textit{Padmanabhan T.}. Structure Formation in the
Universe. - Cambridge University Press, Cambridge, 1993.
\bibitem{NP83}\textit{Narlikar J.V. and Padmanabhan T.}//  Phys.
Rep. - 1983. - \textbf{100}. - P.152.
\bibitem{HE77}\textit{Hawking S.W. and Ellis G.F.R.}. The Large Scale Structure of Space-Time.
 - Cambridge University Press, Cambridge, 1973.
\bibitem{KJ98}\textit{Kiefer K. and Joos E.}, in Quantum Future,
Blanchard P. and Jadczyk A. eds. - Springer-Verlag, Heidelberg,
1998; quant-ph/9803052.
\bibitem{Ki99}\textit{Kiefer K.}, in Lecture Notes in Physics 541: Towards
Quantum Gravity, Kowalski-Glikman J. ed. - Springer-Verlag,
Heidelberg, 2000; gr-qc/9906100.
\bibitem{Zeh99}\textit{Zeh H.D.}. The Physical Basis of the
Direction of Time. - Springer-Verlag, Heidelberg, 1999.
\bibitem{D68}\textit{Dirac P.A.M.}. Lectures on Quantum Mechanics. -
Yeshiva University, New York, 1964.
\bibitem{W67}\textit{DeWitt B.S.}// Phys. Rev.  - 1967. - \textbf{160}. -
P.1113; \textit{Wheeler J.A.}, in Battelle Rencontres, de Witt C.
and Wheeler J.A. eds. - Benjamin, New York, 1968.
\bibitem{E17}\textit{Einstein A.}. Relativity: The Special and the
General Theory. - Crown, New York, 1961.
\bibitem{H17}\textit{Hilbert D}.//  Nachr. Ges. Wiss.
G\"{o}ttingen - 1917. - \textbf{53}. - P.1.
\bibitem{KT91}\textit{Kucha\v{r} K.V. and Torre C.G.}//  Phys. Rev.
- 1991. - \textbf{D43}. - P.419.
\bibitem{BK95}\textit{Brown J.D. and Kucha\v{r} K.V.}//  Phys. Rev.
- 1995. - \textbf{D51}. - P.5600.
\bibitem{BM96}\textit{Brown J.D. and Marolf D.}//  Phys. Rev.
- 1996. - \textbf{D53}. - P.1835; gr-qc/9509026.
\bibitem{ShS04}\textit{Shestakova T.P. and Simeone C.}// Grav. Cosmol. -
2004. - \textbf{10}. - P.161; gr-qc/0409114.
\bibitem{PS01}\textit{Peskin M.E. and Schroeder D.V.}. An Introduction to 
Quantum Field Theory. - Addison-Wesley Publishing Company, Reading, 1995.
\bibitem{D88} \textit{Dolgov A.D., Zeldovich Ya.B., and Sazhin M.V.}.
Kosmologiya rannei vselennoi (Cosmology of the Early Universe).
 - Moscow University, Moscow, 1988.
\bibitem{L90}\textit{Linde A.D.}. Elementary Particle Physics and
Inflationary Cosmology. - Harwood, Chur, 1990.
\bibitem{LR99}\textit{Lyth D.H. and Riotto A.}// Phys. Rep. -
1999. - \textbf{314}. - P.1; hep-ph/9807278.
\bibitem{W72}\textit{Weinberg S.}. Gravitation and Cosmology. - Wiley, New
York, 1972.
\bibitem{V04}\textit{Kuzmichev V.V.}// JINA - Virt. J. Nucl. Astrophys.
- 2004. - \textbf{2}, Issue 26; astro-ph/0407013.
\bibitem{ADM}\textit{Arnowitt R., Deser S., and Misner C.W.} in
Gravitation: An Introduction to Current Research, Witten L. ed. -
Wiley, New York, 1962.
\bibitem{K76}\textit{Kucha\v{r} K.V.}// J. Math. Phys. -
1976. - \textbf{17}. - P.801.
\bibitem{VV04b}\textit{Kuzmichev V.E. and Kuzmichev V.V.}, in
Trends in Dark Matter Research, Blain J.V. ed. - Nova Science
Publishers, Hauppauge, 2005; astro-ph/0405455.
\bibitem{Kof96}\textit{Kofman L.A.}, in Relativistic
Astrophysics: A Conference in Honor of Igor Novikov's 60th
Birthday, Jones B. and Markovic D. eds. - Cambridge University
Press, Cambridge, 1996; astro-ph/9605155.
\bibitem{F76}\textit{Fock V.A.}. Nachala kvantovoi mekhaniki (Foundation
of Quantum Mechanics). - Nauka, Moscow, 1976.
\bibitem{B71}\textit{Baz' A.I., Zel'dovich Ya.B., and Perelomov A.M.}.
Scattering, Reactions, and Decays in Nonrelativistic Quantum
Mechanics. - Israel Program of Sci. Transl., Jerusalem, 1966.
\bibitem{L88}\textit{Landau L.D., Lifshitz E.M.}. The Classical Theory of
Fields. - Pergamon, Oxford, 1975.
\bibitem{M75}\textit{Migdal A.B.}. Kachestvenniye metody v kvantovoi teorii
(Qualitative Methods in Quantum Theory). - Nauka, Moscow, 1975.
\bibitem{D64}\textit{Dicke R.H.}, in Gravitation and Relativity,
Hong-Yee Chiu and Hoffmann W.F. eds. - Benjamin, New York, 1964.
\bibitem{VV04}\textit{Kuzmichev V.E. and Kuzmichev V.V.}, in
Quantum Cosmology Research Trends. Horizons in World Physics,
Volume 246, Reimer A. ed. - Nova Science Publishers, Hauppauge,
2005; astro-ph/0405454.
\bibitem{DD04} \textit{Daly R.A. and Djorgovski S G.}// Astrophys.
J.- 2004. - \textbf{612}. - p.845; astro-ph/0403664.
\bibitem{D79}\textit{Dirac P.A.M.}. The Principles of Quantum Mechanics. -
Clarendon Press, Oxford, 1958.
\bibitem{ZN71} \textit{Zel'dovich Ya.B., Novikov I.D.}. Teoriya tyagoteniya
i evolutsiya zvezd (The Theory of Gravity and Evolution of Stars).
- Nauka, Moscow, 1971.
\bibitem{G03} \textit{Gott J.R. et al.}// astro-ph/0310571.
\bibitem{VV03}\textit{Kuzmichev V.E. and Kuzmichev V.V.}// Ukr. J.
Phys. - 2003. - \textbf{48}. - P.801 ; astro-ph/0301017.

\end{thebibliography}
\end{document}